\documentclass[twocolumn,english,superscriptaddress,floatfix,longbibliography]{revtex4-1}
\usepackage[T1]{fontenc}
\usepackage[latin9]{inputenc}
\usepackage{amsmath}
\usepackage{amssymb}
\usepackage{graphicx}

\makeatletter

\providecommand{\tabularnewline}{\\}

\usepackage{braket}
\usepackage{upgreek}
\usepackage{amssymb}

\makeatother

\usepackage{babel}
\begin{document}
\title{Dynamical breaking of inversion symmetry, strong second harmonic generation,
and ferroelectricity with nonlinear phonons}
\author{Egor I. Kiselev}
\affiliation{Max-Planck-Institut für Physik komplexer Systeme, 01187 Dresden, Germany}
\begin{abstract}
We show how crystalline inversion symmetry can be dynamically broken
by optical phonons with generic, hardening Kerr-like non-linearities.
The symmetry-broken state is reached through a parametric instability
that can be accessed by driving close to half the phonon frequency.
The system then settles to a steady state with inversion-symmetry
breaking phonon trajectories and strong second harmonic generation.
The time averaged positions of the atoms are displaced relative to
equilibrium, resulting in a ferroelectric rectification of the driving
signal. For circularly polarized phonons, complex Lissajous-like trajectories
can be achieved.
\end{abstract}
\maketitle

\paragraph*{Introduction}

The non-equilibrium behavior of condensed matter systems currently
attracts considerable attention due to its potential for on-demand
control over materials \citep{basov2017_on_demand,bloch2022strongly_corr_el_photon,rudner2020band}.
Out-of-equilibrium phonons are of particular interest, as lattice
distortions have an immediate impact on the electronic properties
of solids \citep{forst2011nonlinear_phononics,mankowsky2016nonlinear_phononics_2}.
Nonlinear lattice dynamics can be expoited to enhace and manipulate
superconductivity \citep{mankowsky2014_enhanced_superconductivity,knap2016phonon_superconduct_theory,babadi2017phonon_superconduct_theory,cavalleri2018photo_induced_supercond,liu2020_res_phonon_light_induced_superc},
magnetism \citep{fechner2018magnetophononics_review,afanasiev2021ultrafast_control_magnetic_phonons,disa2023_phonon_ferromag_induce,luo2025terahertz_control_magno_phononics}
and other states of matter \citep{nova2019phonon_ferroelectricity,ning2023_phonon_induced_hidden_quadrupolar,kaplan2025_spatiotemporal_phonons,kaplan2025spatiotemporal_phonons_first_principles}.
Other effects predicted for driven, nonlinear collective modes include
non-trivial steady states \citep{kiselev2024MFPD_plasmons_nat,kiselev2024_PRB_MFPD,kiselev2025exciting,wanic2025entanglement,kaplan2025_spatiotemporal_phonons,kaplan2025spatiotemporal_phonons_first_principles,kaplan2025optically_induced_goldstone_faraday},
chaos \citep{wanic2024magnetoelectric}, and parametric amplification
\citep{cartella2018parametric,buzzi2021higgs_lasing,michael2024photonic}. 

Here, we show how IR-active phonons oscillating in anharmonic interatomic
potentials can be driven in unconventional ways to break underlying
crystal symmetries and create symmetry-forbidden electromagnetic responses.
We focus on inversion symmetry, which is known to prohibit second
harmonic generation (SHG) and rectification \citep{Boyd_nonlinear_optics}.
We demonstrate that, for a very generic hardening Kerr-like nonlinearity,
this rule can be circumvented by a taylored driving protocol, resulting
in a symmetry-breaking steady state with strong phononic SHG and DC
responses. We stress that no double well-structure with metastable
ferroelectric states, as e.g. exploited in Ref. \citep{li2019_driven_metastable_ferroelectricity},
is required.

The DC response leads to a static displacement of atoms from their
equilibrium positions, generating a constant-in-time dipolar electric
field. This behavior is analogous to that of an equilibrium ferroelectric,
where -- below a critical temperature -- dipolar fields form due
to spontaneous, symmetry-breaking lattice deformations. In our case,
the static dipolar response is triggered by a dynamical breaking of
inversion symmetry, and we argue that the predicted effect can be
used to create an on-demand ferroelectric. In particular, we show
that the non-equilibrium ferroelectric response exhibits hysteretic
behavior, which underlines that the effect is not merely a nonlinear
piezoelectric response \citep{disa2020polarizing_symmetry_breaking_light_nonlinearity},
but instead a symmetry-breaking out-of-equilibrium phase. We show
that this phase is robust to noise, and is accessible at reasonable
field strenghts.

We then explore the above effects in a system of circularly polarized
phonons \citep{juraschek2025chiral_phonons_rev}, which are sometimes
referred to as chiral phonons \citep{kahana2024_chiral_rectification},
and have been used to manipulate magnetic states. Here, the strong
SHG leads to Lissajous-like, non-inversion-symmetric phonon trajectories.
Such trajectories create structured magnetic fields influencing the
dynamics of electrons and spins on a microscopic level \citep{luo2023large_magnetic_chiral_phonons,juraschek2022giant_magn_field_chiral_phonons,xiong2022effective_magn_field_chiral},
and provide an interesting way of engineering electronic behavior
\citep{yaniv2025multicolor}.

\paragraph{Symmetry-breaking instability and second harmonic generation}

\begin{figure*}
\begin{centering}
\includegraphics[scale=0.32]{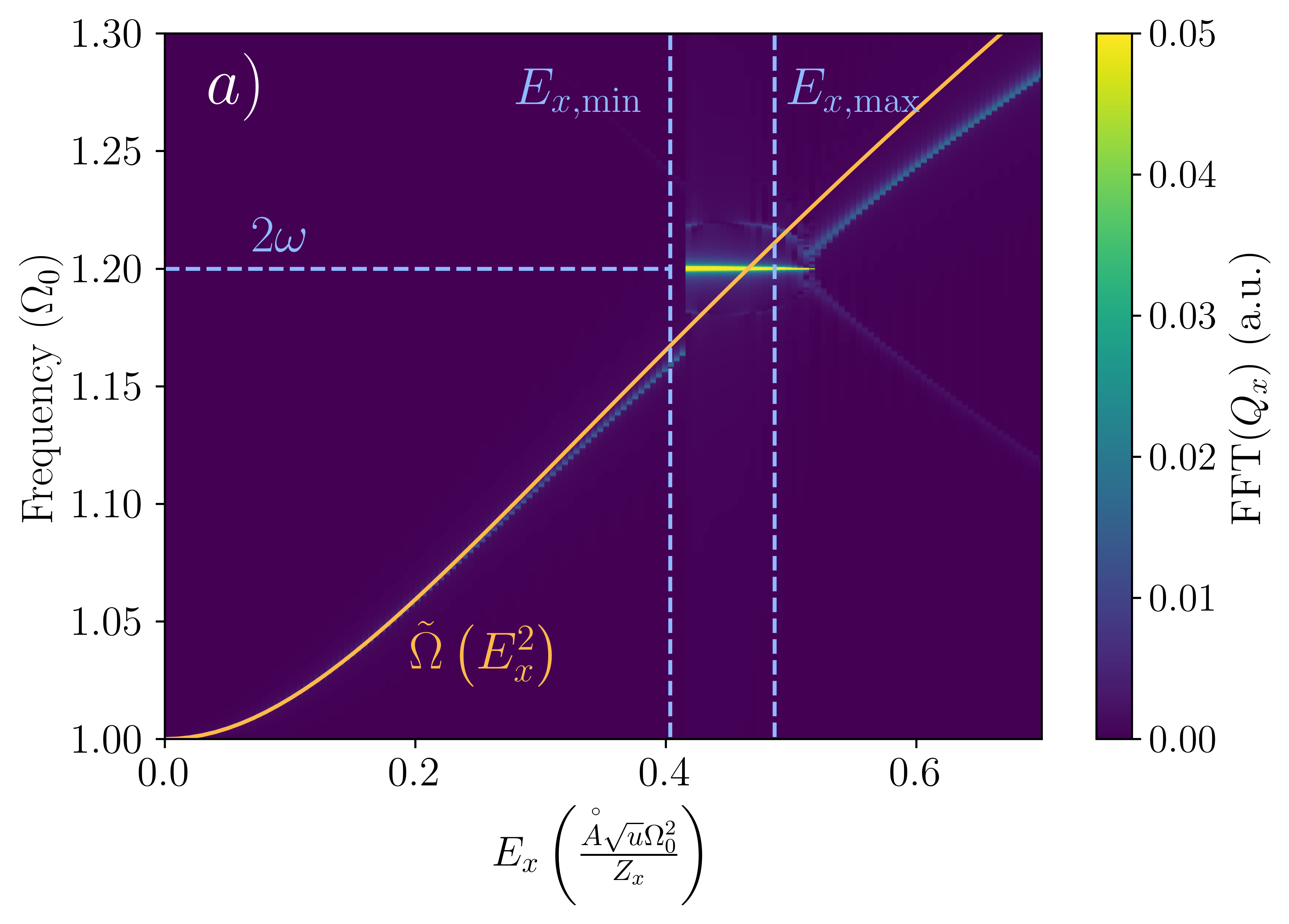}\includegraphics[scale=0.32]{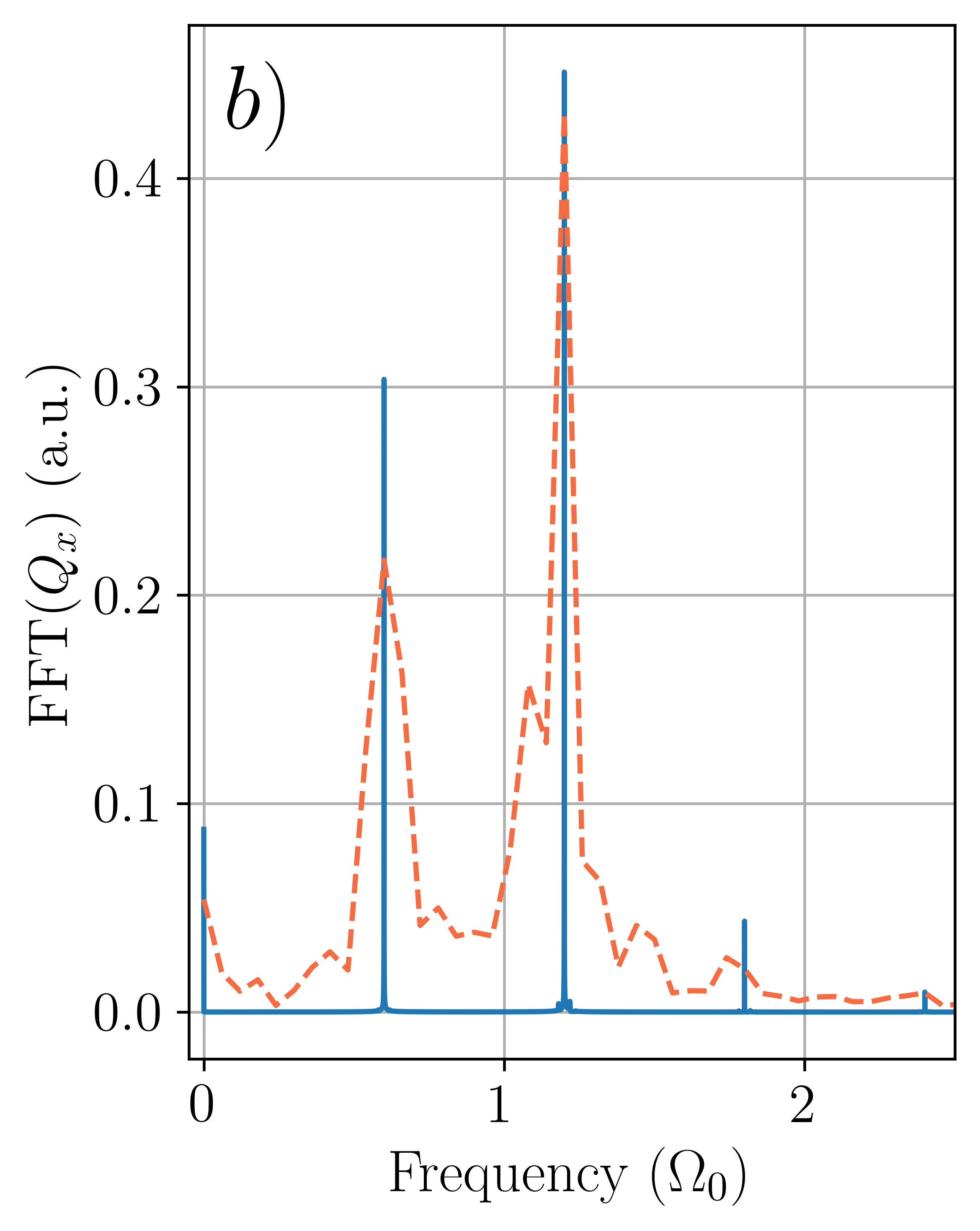}\includegraphics[scale=0.32]{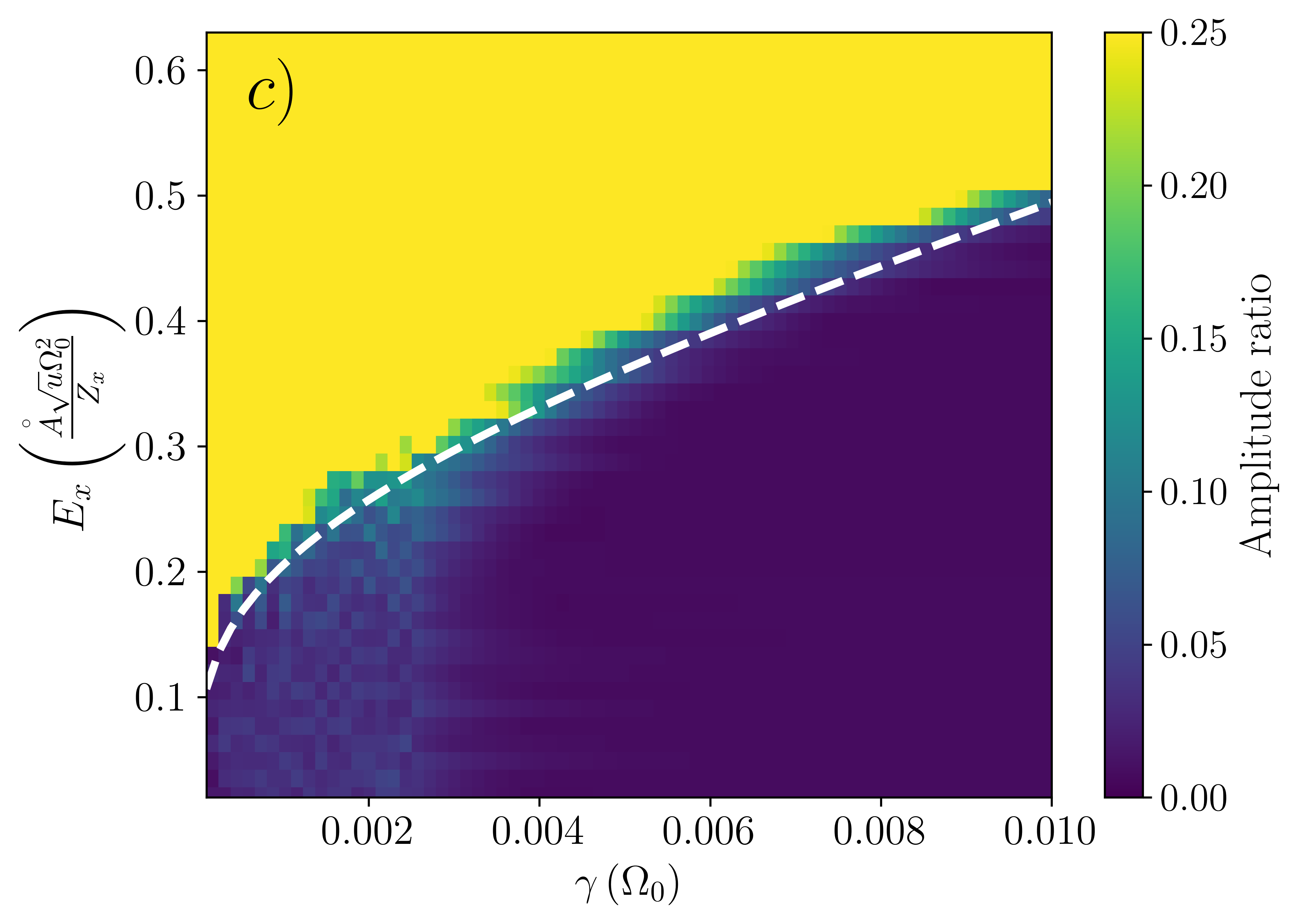}\includegraphics[scale=0.32]{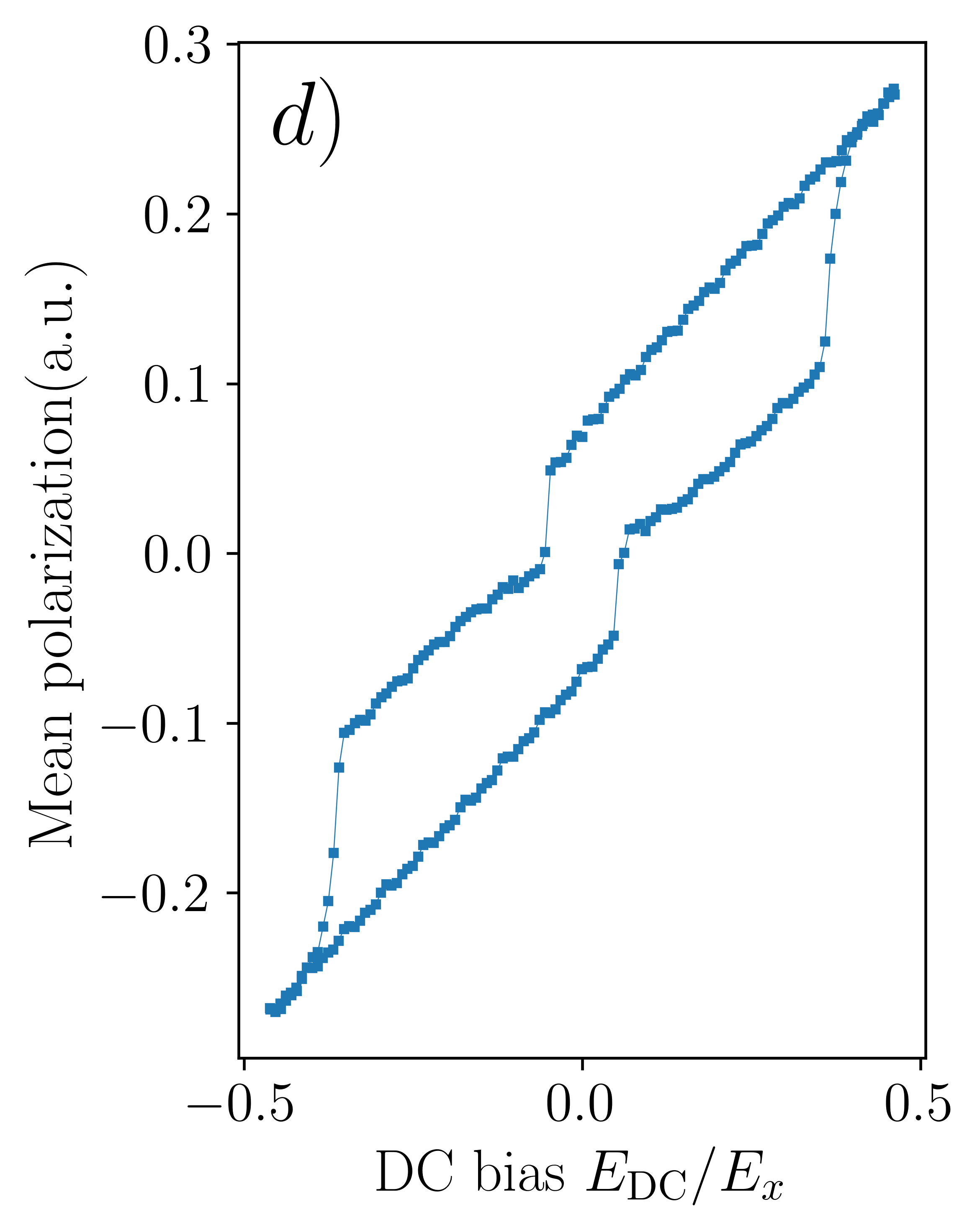}
\par\end{centering}
\centering{}\caption{a) Phonons {[}Eq. (\ref{eq:Phonon_H}){]} driven with a linearly polarized
electric field $E_{x}\cos\left(\omega t\right)$ oscillating at frequency
$\omega=0.6\Omega_{0}$. The effective phonon frequency $\tilde{\Omega}_{0}\left(E_{x}\right)$,
given in Eq. (\ref{eq:shifting_frequency}), exhibits a blue shift
as $E_{x}$ is increased. Around $\tilde{\Omega}_{0}\left(E_{x}\right)\approx2\omega$,
the system enters the symmetry breaking state with strong second harmonic
generation. In the symmetry-breaking regime, the resonance curve $\tilde{\Omega}_{0}\left(E_{x}\right)$
is interrupted. A small damping of $\gamma=0.001$ was chosen to have
a large instability interval for clarity. b) Spectrum of $Q_{x}\left(t\right)$
in the symmetry breaking state. c) Amplitude ratio of second and first
harmonics across the symmetry breaking transition for different dampings
$\gamma$. d) Hysteresis of the mean polarization $p_{x}=Z_{x}F_{x,\text{0}}=\left(2\pi/\omega\right)\int_{0}^{2\pi/\omega}dt\,Q_{x}\left(t\right)$
when a static bias field $E_{\mathrm{\mathrm{DC}}}$ is swept through
positive and negative values in a triangle-wave-like manner.\label{the_fig}}
\end{figure*}

To show how the symmetry breaking instability emerges, we first focus
on a single IR-active phonon mode $Q_{x}\left(t\right)$. For a generic,
inversion symmetric system, the dynamics of $Q_{x}\left(t\right)$
is governed by the Hamiltonian
\begin{equation}
H=\frac{1}{2}P_{x}^{2}+\frac{\Omega_{0}^{2}}{2}Q_{x}^{2}+\frac{\beta}{4}Q_{x}^{4}+V_{\mathrm{l-m}}.\label{eq:1D_Hamiltonian}
\end{equation}
Here, $Q_{x}$ is given in units of $\text{Å}\sqrt{u}$, where $u$
is the atomic mass unit, $\Omega_{0}$ is the resonance frequency
of the phonon, and $\beta>0$ controls the strength of the nonlinearity.
Notice that, for positive $\beta$, unlike for $\beta<0$, the lattice
potential has no metastable minima at $Q_{x}\neq0$. This underlines
the truly dynamical nature of the symmetry breaking described in the
following. Finally, $V_{\mathrm{l-m}}$ is a the dipolar coupling
between the phonon and an electromagnetic field:
\begin{equation}
V_{\mathrm{l-m}}=-\mathbf{E}\cdot\mathbf{p}_{x},
\end{equation}
where the electric dipole moment of the phonon is given by $\mathbf{p}_{x}=\mathbf{Z}_{x}Q_{x}$,
with effective charge $\mathbf{Z}_{x}$. We assume $\mathbf{Z}_{x}\propto\hat{\mathbf{e}}_{x}$.
Then, for circularly polarized light, the phonon equation of motion
reads
\begin{align}
\ddot{Q}_{x}+2\gamma\dot{Q}_{x}+\Omega_{0}^{2}Q_{x}+\beta Q_{x}^{3} & =Z_{x}E_{x}\cos\omega t\label{eq:EoM_single_mode}
\end{align}
where we included a damping term with damping rate $\gamma$. 

It is often assumed that the third order nonlinearity of Eq. (\ref{eq:EoM_single_mode})
does not generate even harmonics of the driving frequency $\omega$.
Indeed, shifting the time coordinate according to $t\rightarrow t+\frac{\pi}{\omega}$
changes the sign of the right hand side, which seems to enforce the
symmetry $Q_{x}\left(t+\frac{\pi}{\omega}\right)=-Q_{x}\left(t\right)$,
excluding any response at frequencies $2n\omega$, where $n\in\mathbb{N}$,
including zero. We now show, that this symmetry, stemming from the
inversion symmetry of the Hamiltonian (\ref{eq:1D_Hamiltonian}),
can be dynamically broken. 

It is useful to divide $Q_{x}\left(t\right)$ into parts composed
of odd and even harmonics:
\begin{equation}
Q_{x}\left(t\right)=Q_{x,\mathrm{odd}}\left(t\right)+Q_{x,\mathrm{even}}\left(t\right),\label{eq:decomposition_Q}
\end{equation}
which are, respectively, antisymmetric and symmetric under a time
translation by half the oscillation period of the electromagnetic
field:
\begin{align}
Q_{x,\mathrm{odd}}\left(t+\frac{\pi}{\omega}\right) & =-Q_{x,\mathrm{odd}}\left(t\right),\nonumber \\
Q_{x,\mathrm{even}}\left(t+\frac{\pi}{\omega}\right) & =Q_{x,\mathrm{even}}\left(t\right).\label{eq:odd_even_req}
\end{align}

We now show how even harmonics ($Q_{i,\mathrm{even}}$) can be created
via a parametric instability. Let us first study the onset of this
instability. Using the decomposition of Eq. (\ref{eq:decomposition_Q}),
we can separate Eq. (\ref{eq:Eqs_of_motion_circular_1}) into cupled
equations for $Q_{x,\mathrm{odd}}$ and $Q_{x,\mathrm{even}}$. We
use that, at the onset of the instability, $Q_{i,\mathrm{even}}$
will be very small, such that $\left|Q_{i,\mathrm{even}}\right|\ll\left|Q_{i,\mathrm{odd}}\right|$.
Then, the equation for the odd part, neglecting contributions stemming
from $Q_{i,\mathrm{even}}$, reads
\begin{equation}
\ddot{Q}_{x,\mathrm{odd}}+2\gamma\dot{Q}_{x,\mathrm{odd}}+\Omega_{0}^{2}Q_{x,\mathrm{odd}}+\alpha Q_{x,\mathrm{odd}}^{3}=Z_{x}E_{x}\cos\left(\omega t\right).
\end{equation}
For our purposes it is sufficient to approximate the response $Q_{x,\mathrm{odd}}$
with the fundamental harmonic and write 
\begin{equation}
Q_{x,\mathrm{odd}}\approx F_{x,1}\left(E_{x}\right)\cos\left(\omega t+\varphi_{x}\right).\label{eq:duffing_1harm_response}
\end{equation}
$F_{x,1}\left(E_{x}\right)$ is then found by inverting the amplitude
equation
\begin{equation}
F_{x,1}^{2}\left[4\gamma^{2}\omega^{2}+\left(\left(\omega^{2}-\Omega_{0}^{2}\right)-\frac{3}{4}\beta F_{x,1}^{2}\right)^{2}\right]=Z_{x}^{2}E_{x}^{2}.\label{eq:duffing_amplitude_eq}
\end{equation}

For the even component $Q_{x,\mathrm{even}}$, we find the Mathieu
equation
\begin{align}
 & \ddot{Q}_{x,\mathrm{even}}+2\gamma\dot{Q}_{x,\mathrm{even}}\nonumber \\
 & +\tilde{\Omega}_{0}^{2}\left(E_{x}\right)\left[1+h\left(E_{x}\right)\cos\left(2\omega t+2\varphi_{x}\right)\right]Q_{x,\mathrm{even}}=0\label{eq:prametric_osc_even}
\end{align}
where $\tilde{\Omega}_{0}\left(E_{x}\right)$ is an effective, amplitude
dependent resonance frequency {[}see Fig. \ref{the_fig}a){]} given
by 
\begin{equation}
\tilde{\Omega}_{0}\left(E_{x}\right)=\Omega_{0}\sqrt{1+\frac{3\beta}{2\Omega_{0}^{2}}F_{x,1}^{2}\left(E_{x}\right)},\label{eq:shifting_frequency}
\end{equation}
and $h\left(E_{x}\right)=3\alpha F_{x,1}^{2}\left(E_{x}\right)/\left[2\tilde{\Omega}_{0}^{2}\left(E_{x}\right)\right]$.
We used Eq. (\ref{eq:duffing_1harm_response}) to approximate $Q_{x,\mathrm{odd}}^{2}$.
It is then the constant-in-time part of $Q_{x,\mathrm{odd}}^{2}$
that modifies the resonance frequency of the mode and leads to a blue
shift, while the oscillating part of $Q_{x,\mathrm{odd}}^{2}$ acts
as a parametric driving for $Q_{x,\mathrm{even}}$. The Mathieu equation
(\ref{eq:param_even}) is known to exhibit parametric instabilities
for $\tilde{\Omega}_{0}\left(E_{x}\right)=n\omega$, with $n$ a positive
integer \citep{Landau_Lifshitz_Mechanics}. However, $Q_{x,\mathrm{even}}$has
to obey Eq. (\ref{eq:odd_even_req}), which excludes the $n=1$ resonance.
The $n=2$ resonance, however, is allowed, and leads to the symmetry-breaking
instability we want to study. Here $\tilde{\Omega}_{0}\left(E_{x}\right)=2\omega$,
such that for driving slightly above half the original resonance frequency
of $\Omega_{0}$, we expect a response at $\tilde{\Omega}_{0}\left(E_{x}\right)$
-- i.e., we expect strong SHG.

As is typical for parametric resonances, the instability occurs in
a small frequency window where for $\Delta=2\omega-\tilde{\Omega}_{0}$
holds (see e.g. \citep{turyn1993damped_Mathieu}, p. 394 \citep{Note2})
\begin{align}
 & \frac{\tilde{\Omega}_{0}}{24}\left(3\left(\frac{4\gamma^{2}}{\tilde{\Omega}_{0}^{2}}-\sqrt{h^{4}-\frac{64\gamma^{2}}{\tilde{\Omega}_{0}^{2}}}\right)+2h^{2}\right)<\Delta\nonumber \\
 & <\frac{1}{24}\tilde{\Omega}_{0}\left(3\left(\frac{4\gamma^{2}}{\tilde{\Omega}_{0}^{2}}+\sqrt{h^{4}-\frac{64\gamma^{2}}{\tilde{\Omega}_{0}^{2}}}\right)+2h^{2}\right).\label{eq:instability_interval}
\end{align}
Here, and in what follows, we omit writing out the $E_{x}$-dependence
of $F_{x,1}$, $\tilde{\Omega}_{0}$, $h$ and $\Delta$ explicitly,
except when it is needed for clarity. The blue shift and the instability
window are shown with the results of a numerical simulation in Fig.
\ref{the_fig} a) (see Appendix\ref{subsec:initial_conditions} for
a note on the initial conditions used). We stress that, while in our
analytical calculation we approximated the odd response by a single
harmonic and linearized in $Q_{x,\mathrm{even}}$, we used the full
Eq. (\ref{eq:EoM_single_mode}) for the simulation. As can be expected
there are discrepancies due to the approximations made, however, the
instability window is clearly identifiable.

To overcome damping effects, a minimal driving amplitude is required.
This threshold amplitude $E_{x,*}$ can be calculated by setting $\Delta=0$.
To leading order in $\gamma/\Omega$, we find $h\left(E_{x,*}\right)=\sqrt{8\gamma/\Omega_{0}}$.
This expression can be inverted for $E_{x}$ using Eq. (\ref{eq:duffing_amplitude_eq}).
For small damping, the threshold electric field amplitude is then
given by

\begin{equation}
E_{x,*}\approx\frac{\sqrt{3}\Omega^{3}}{2^{3/4}\sqrt{\beta}Z_{x}}\left(\frac{\gamma}{\Omega_{0}}\right)^{1/4}.\label{eq:threshold_field}
\end{equation}
Notice that $\Omega_{0}$ is the bare resonance frequency, but the
blue-shift has been taken into account in the derivation of Eq. (\ref{eq:threshold_field}).
As can be expected, $E_{x,*}$ is lowered by a strong nonlinearity
$\beta$ and increased by a larger $\gamma$. The result of Eq. (\ref{eq:threshold_field})
is confirmed by numerical simulations as shown in Fig. \ref{the_fig}
c). The threshold fields for different (bare) phonon resonance frequencies
$\Omega_{0}$ are given in Table \ref{tab:Critical-driving-fields}.
Since $E_{x,*}\sim\Omega_{0}^{3}$ holds, much stronger amplitudes
are required at higher frequencies. The amplitudes are, however, in
an experimentally accessible range.
\begin{table}
\begin{centering}
\begin{tabular}{|c|c|c|c|}
\hline 
$\Omega_{0}/2\pi$ & $1\,\mathrm{THz}$ & $3\,\mathrm{THz}$ & $5\,\mathrm{THz}$\tabularnewline
\hline 
\hline 
$E_{x,*}$ & $1.5\cdot10^{6}\,\mathrm{V/m}$ & $4.1\cdot10^{7}\,\mathrm{V/m}$ & $1.9\cdot10^{8}\,\mathrm{V/m}$\tabularnewline
\hline 
\end{tabular}\caption{Critical driving fields $E_{x,*}$ according to Eq. (\ref{eq:threshold_field})
for the parameters $\gamma/\Omega=0.05$, $Z_{x}=\mathrm{e}/\sqrt{\mathrm{u}}$,
and $\beta=1\,\mathrm{eV}\mathrm{u^{-2}}\text{Å}^{-4}$ \citep{vonHoegen2018probing}.\label{tab:Critical-driving-fields}}
\par\end{centering}
\end{table}

\paragraph{Symmetry breaking steady-state and pulsed excitation}

The above analysis demonstrates that the usual solution, consisting
of a strong response at the fundamental harmonic and perturbative
corrections at odd harmonics, is unstable if driving frequency and
strength are tuned such that the condition of Eq. (\ref{eq:instability_interval})
at $2\omega\approx\tilde{\Omega}_{0}\left(E_{x}\right)$ is fulfilled.
This instability is associated with strong second harmonic generation
due to the parametric resonance in Eq. (\ref{eq:param_even}). Investigating
the dynamics of the nonlinear phonon mode in this regime, we find
that the system converges to a stable steady-state which breaks the
inversion symmetry of the system. Besides a strong second harmonic,
this symmetry-breaking is characterized by a relatively large DC offset.
The spectrum of $Q_{x}$ in the steady state, obtained by solving
Eq. (\ref{eq:EoM_single_mode}) numerically, is shown in Fig. \ref{the_fig}
b). We note in passing, that the instability and steady state studied
here have been described in nonlinear systems literature, although
the only extensive study, to our knowledge, is presented in Ref. \citep{ys1991_duffing_symm_breaking_SHG}. 

For low frequency phonons the instability is accessible already at
relatively small driving fields (see Table\ref{tab:Critical-driving-fields}),
which might be produced by continuous wave lasers. However, reaching
the symmetry breaking with pulsed signals is also possible. We performed
a series of numerical experiments using the same parameters as in
the continuous wave case, and find that short pulses with as few as
five oscillation cycles (see Appendix \ref{subsec:Achieving-symmetry-breaking})
are sufficient to observe both second harmonic generation and a DC
offset. However, flat-top pulses must be used, because, due to the
nonlinear blue-shift, the frequency and amplitude of the drive must
be tuned toghether. The dashed lines in Fig.\ref{the_fig} b) show
the response spectrum obtained for a flat top pulse of five optical
cycles. The finite duration of the pulse introduces spectral broadening,
however, the SHG and the DC offset are clearly visible. 

Before extending our results to a system of chiral phonons, we investigate
how the DC component of the steady-state at $2\omega\approx\tilde{\Omega}_{0}\left(E_{x}\right)$
leads to a ferroelectric response, and show how auxiliary phonon modes
can be exploited to trigger the symmetry breaking instability resonantly.

\paragraph{Ferroelectricity, hysteresis, stability to noise}

We now show that the symmetry-breaking steady-state outlined above,
necessarily implies the presence of a static displacement of the atoms
from their equilibrium positions. To see this, we average Eq. (\ref{eq:EoM_single_mode})
over one period of the drive. Writing $Q_{x}\left(t\right)=\sum_{a}F_{x,a}\cos\left(a\omega t+\varphi_{x,a}\right)$,
and truncating the series at $a=2$, we find
\begin{align}
 & \omega_{0}^{2}F_{x,0}+\beta\left[\frac{3}{2}\left(F_{x,1}^{2}+F_{x,2}^{2}\right)F_{x,0}\right.\nonumber \\
 & \quad+\left.\frac{3}{4}F_{x,1}^{2}F_{x,2}\cos\left(2\varphi_{x,1}-\varphi_{x,2}\right)+F_{x,0}^{3}\right].
\end{align}
This equation has one non-trivial, real solution for $F_{x,0}$. To
leading order in $F_{x,1}$ and $F_{x,2}$, it reads
\begin{equation}
F_{x,0}=-\frac{3\beta}{4\omega_{0}^{2}}F_{x,1}^{2}F_{x,2},
\end{equation}
showing that any response at the second harmonic is accompanied by
a DC offset. Being third order in the first and second harmonic amplitudes,
we expect the DC offset to be smaller in magnitude, it can however,
still be sizable {[}see Fig. (\ref{the_fig}) b){]}. 

We conclude that although the symmetry breaking instability is triggered
by an oscillating driving field, inversion symmetry is statically
broken. This results in constant-in-time electric fields produced
by the dipoles $\mathbf{p}=\mathbf{Z}_{x}F_{x,0}$, where $\mathbf{Z}_{x}$
is the effective electric charge of the phonon mode in question. The
driving signal is thus rectified and triggers a ferroelectric response.

This dynamically triggered ferroelectric response shares significant
similarities with equilibrium ferroelectrics, showing that it can
be used as a road to on-demand light driven ferroelectricity. To substantiate
this claim, we demonstrate that hysteresis can be observed when the
phonon mode is in the symmetry-breaking state. To this end we add
a static component $E_{\mathrm{DC}}$ to the driving field in Eq.
(\ref{eq:EoM_single_mode}) and sweep it through positive and negative
values in a triangle-wave-like manner. We numerically integrate Eq.
(\ref{eq:EoM_single_mode}) while slowly changing $E_{\mathrm{DC}}$.
Fig.\ref{the_fig}d) shows the mean polarization
\begin{equation}
p_{x}=Z_{x}F_{x,\text{0}}=\left(2\pi/\omega\right)\int_{0}^{2\pi/\omega}dt\,Q_{x}\left(t\right),
\end{equation}
which exhibits a hysteresis. Interestingly, the hysteresis curve is
pinched -- also a feature known from equilibrium ferroelectrics \citep{xu2016pinched_hysteresis_ferroelectrics}. 

\begin{figure}
\begin{centering}
\includegraphics[width=0.9\columnwidth]{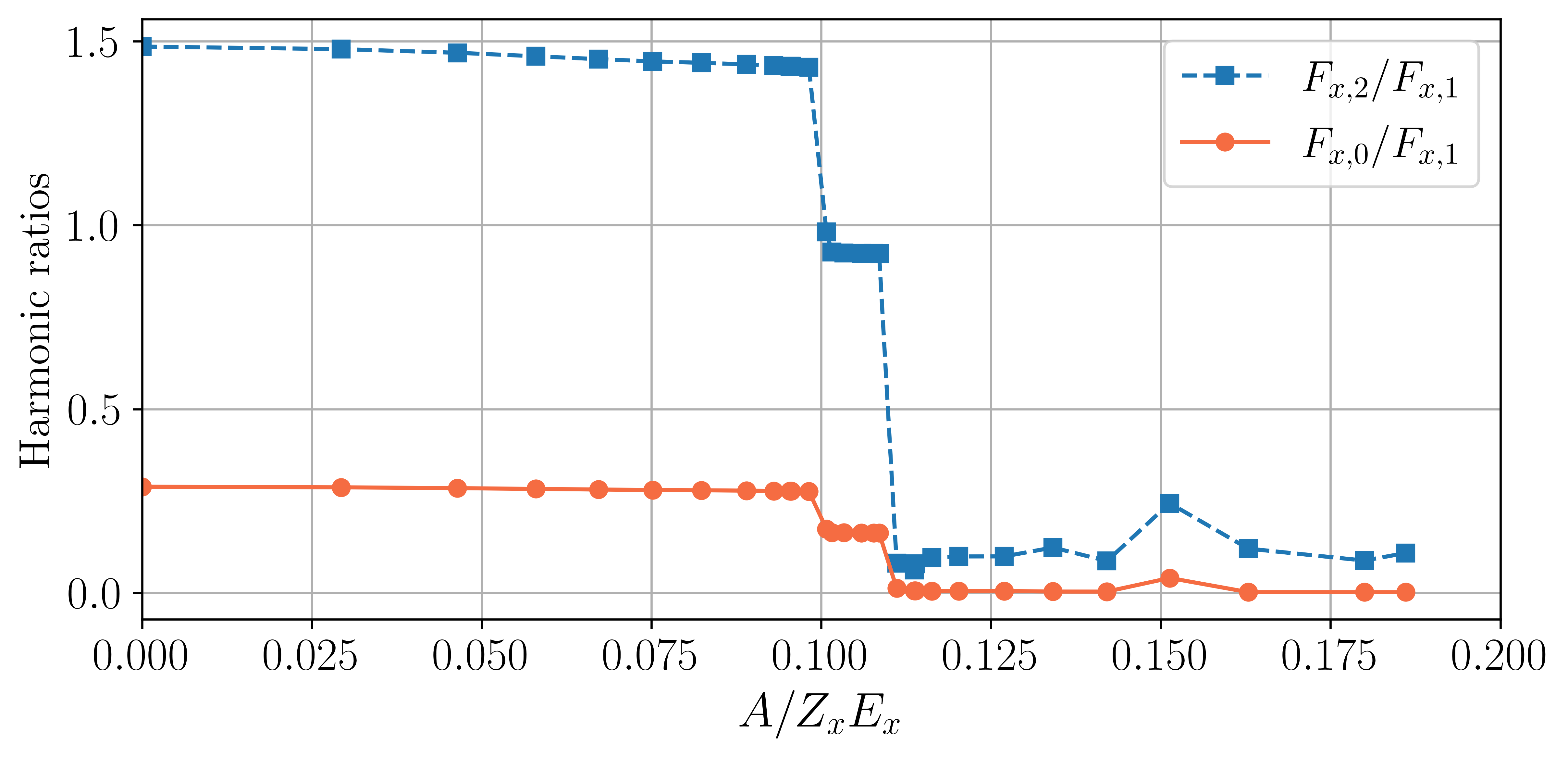}
\par\end{centering}
\centering{}\caption{Collapse of the symmetry-breaking steady state under the influence
of noise. In the case of thermal noise and $\Omega_{0}/2\pi=3\,\mathrm{THz}$,
the critical noise strength $A\approx0.1Z_{x}E_{x}$ corresponds to
a temperature of $\sim700\,\mathrm{K}$. The same parameters as in
Fig.\ref{the_fig} were used. \label{fig:noise_robustness}}
\end{figure}

\begin{figure}
\begin{centering}
\includegraphics[width=0.9\columnwidth]{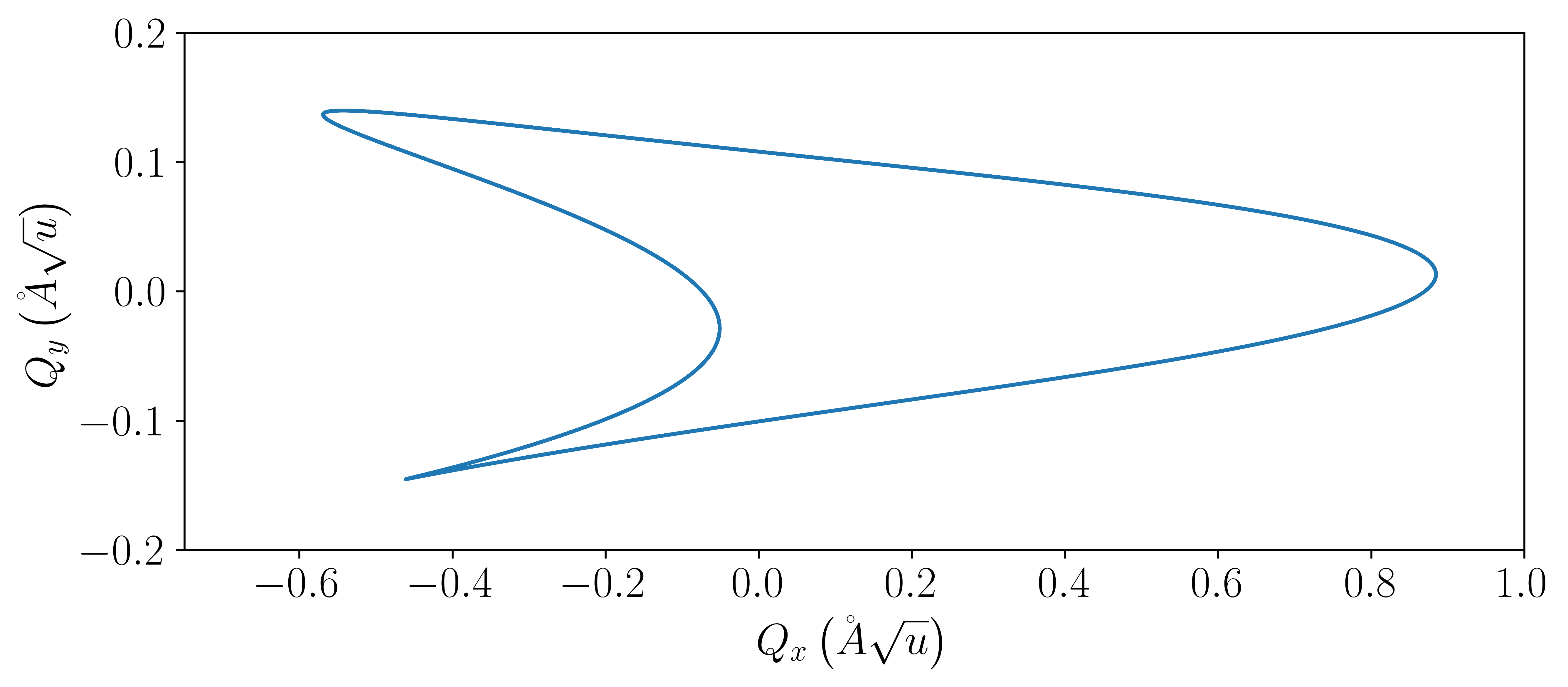}
\par\end{centering}
\centering{}\caption{The Lissajous trajectory of phonon coordinates $Q_{x}\left(t\right)$
and $Q_{y}\left(t\right)$ when driven into the symmetry-breaking
state using elliptically polarized light with $\mathbf{E}=E_{x}\left[\cos\left(\omega t\right),0.25\sin\left(\omega t\right)\right]$.
The inversion symmetry of Eq. (\ref{eq:Phonon_H}) is broken dynamically.\label{fig:lissajous}}
\end{figure}

Finally we want to investigate the sensitivity of the symmetry-breaking
steady state to noise. First, a simple estimation can be made. In
the steady state, we find $\sqrt{\left\langle Q^{2}\right\rangle }\sim1\text{Å}\sqrt{\mathrm{u}}$
(see e.g. Fig. \ref{fig:lissajous}). We can expect that the steady
state breaks down when, roughly, $\Omega_{0}^{2}\left\langle Q^{2}\right\rangle \approx k_{B}T$,
yielding a threshold temperature of $\sim430\,\mathrm{K}$ for the
$3\,$THz mode or $\sim1200\,\mathrm{K}$ for the $5\,$THz mode.
To substantiate this estimate, we perform a simulation of the equation
of motion (\ref{eq:EoM_single_mode}) with an additional Langevin
term $A\eta\left(t\right)$, where $\eta\left(t\right)$ normally
distributed noise that is uncorrelated on physically relevant timescales.
We indeed find that the steady state collapses above a threshold noise
amplitude (see Fig. \ref{fig:noise_robustness} and Appendix\ref{subsec:Noise-robustness}).
Interestingly, the collapse seems to proceed in steps -- an effect
that could be related to the pinched hysteresis of Fig.\ref{the_fig}d).
Using the parameters $\gamma=0.1$, $\Omega_{0}/2\pi=3\,\mathrm{THz}$
and $E_{x,*}=4.1\cdot10^{7}\,\mathrm{V/m}$, we find that this critical
noise amplitude corresponds to a temperature of $\sim700\,\mathrm{K}$.
We thus conclude that the symmetry breaking state is reasonably robust
to perturbations by noise in general, and thermal effects in particular.

\paragraph{Exploiting auxiliary IR- and Raman-active modes}

Auxiliary modes can be used to further simplify reaching the symmetry-breaking
steady state. Infrared-active phonons with frequencies $\omega_{A}$
close to $\Omega_{0}/2$ can be exploited to resonantly enhance the
otherwise off-resonant driving: consider a phonon mode $Q_{A}$ with
a coupling term 
\begin{equation}
H_{qQ}=\lambda Q_{x}Q_{A}.
\end{equation}
This coupling preserves the original inversion symmetry of the system
and leads to $\lambda Q_{A}$ taking over the role of the electric
field in Eq. (\ref{eq:Eqs_of_motion_circular_1}). The mode $q_{A}$
can be pumped resonantly, such that the energy of the drive is stored
over $\sim\omega_{A}/\gamma_{A}$ cycles, resulting in a large amplitude
oscillation of $Q_{A}$. Due to the nonlinear blue-shift of $\tilde{\Omega}_{0}$
{[}Eq. (\ref{eq:shifting_frequency}){]}, the frequency of the auxiliary
phonon $\omega_{A}$ can be adjusted by choosing a suitable driving
amplitude, such that it exactly hits $\omega_{A}=\tilde{\Omega}_{0}/2$.

Raman-active modes can be used in a similar way. Here the lowest order,
inversion symmetric coupling Hamiltonian is given by
\begin{equation}
H=AQ_{R}Q_{x}^{2}.\label{eq:Raman_coupling_H}
\end{equation}
$Q_{R}$ is an auxiliary Raman mode. As shown in Appendix \ref{subsec:Auxiliary-resonances-and},
including the coupling of Eq. (\ref{eq:Raman_coupling_H}) again leads
to an equation of motion for $Q_{x}$ that can be mapped onto the
original Eq. (\ref{eq:Eqs_of_motion_circular_1}). Thus Raman active
modes can, too, be used to drive the system into the symmetry-breaking
state. This might me advantageous, since driving these modes allows
for greater optical penetration depths compared to infrared-active
modes.

\paragraph{Nonlinear, circularly polarized phonons}

We now investigate the effects of dynamical symmetry breaking on circularly
polarized phonons. While sometimes such phonons are referred to as
chiral \citep{kahana2024_chiral_rectification}, this terminology
has been questioned \citep{juraschek2025chiral_phonons_rev}. We will
discuss both, degenerate circularly polarized optical phonons \citep{kahana2024_chiral_rectification,cheng2020_dirac_semi_phonon_zeeman,mustafa2025origin_phonon_zeeman_mos2},
as well as phonons with split frequencies for right- and left-handed
motion \citep{zhang2015chiral_phonons_hexagonal,zhu2018observation_chiral_phonons,ishito2023truly_chiral_phonons,ueda2023chiral_phonons_xrays}. 

A simple model for nonlinear, degenerate, circularly polarized phonons
is given by the Hamiltonian \citep{kahana2024_chiral_rectification}
\begin{equation}
H=\frac{P_{x}^{2}}{2}+\frac{P_{y}^{2}}{2}+\frac{\Omega_{0}^{2}}{2}\left(Q_{x}^{2}+Q_{y}^{2}\right)+\frac{\beta}{4}\left(Q_{x}^{2}+Q_{y}^{2}\right)^{2}+V_{\mathrm{l-m}}.\label{eq:Phonon_H}
\end{equation}
Here, $Q_{x}$ and $Q_{y}$ are the coordinates of two orthogonal
phonon modes. The light-matter coupling $V_{\mathrm{l-m}}$ is given
by
\begin{equation}
V_{\mathrm{l-m}}=-\mathbf{E}\cdot\left(\mathbf{p}_{x}+\mathbf{p}_{y}\right),
\end{equation}
where $\mathbf{p}_{n}=\mathbf{Z}_{n}Q_{n}$ are the electric dipole
moments of the phonon components, with effective charges $\mathbf{Z}_{n}$.
For simplicity, we assume $\mathbf{Z}_{n}\propto\hat{\mathbf{e}}_{n}$.
The equations of motion for elliptically polarized light are given
in Appendix\ref{subsec:Collective-instability-of}, where we added
damping at rate $\gamma$. 

Solving the equations of motion (\ref{eq:Eqs_of_motion_circular_1})
numerically, we observe that the instability intervals are larger
for elliptical polarization with $E_{x}\neq E_{y}$. A stability analysis
for the system following Ref. \citep{Landau_Lifshitz_Mechanics},
for which we refer to Appendix\ref{subsec:Collective-instability-of},
rationalizes this observation.

The result of a numerical simulation of the phonon trajectory in the
symmetry-breaking state is shown in in Fig. \ref{fig:lissajous} for
driving with elliptically polarized light where $E_{y}=0.25E_{x}$.
The resulting Lissajous trajectory breaks the inversion symmetry of
the Hamiltonian (\ref{eq:Phonon_H}), due to the large second harmonic
component of $Q_{x}\left(t\right)$.

We note that beyond the instability at $\omega\approx\Omega_{0}/2$,
higher order, inversion symmetry breaking instabilities at frequencies
$\omega\approx\Omega_{0}/n$, where $n$ is an even number can be
induced. These generate more complex Lissajous-figures, however the
required threshold driving amplitudes grow according to $E_{x,*}\sim\gamma^{1/2n}$
\citep{Landau_Lifshitz_Mechanics}.

Finally, we investigate dynamical symmetry breaking for non-degenerate
circularly polarized phonons \citep{zhang2015chiral_phonons_hexagonal}
(also called ``false chiral phonons'' \citep{juraschek2025chiral_phonons_rev}).
Such phonons can appear in systems with broken time-reversal symmetry
\citep{bonini2023CrI3_chiral_phonons} and have been observed in monolayer
TMDs \citep{zhu2018observation_chiral_phonons}. A toy-model with
split frequencies for phonons of opposite chiralities, is obtained
by substituting $P_{i}\rightarrow P_{i}-\kappa A_{i}$ in the Hamiltonian
of Eq. (\ref{eq:Phonon_H}) \citep{chen2019chiral_chen_review}. Here
$\mathbf{A}=B_{\mathrm{eff}}\left[-Q_{y},Q_{x},0\right]$ takes the
role of an effective magnetic vector potential. To linear order in
$\kappa$, the above substitution is equivalent to adding the term
$\kappa\mathbf{B}_{\mathrm{eff}}\cdot\mathbf{L}$ to the Hamiltonian
(\ref{eq:Phonon_H}), where $\mathbf{L}=\left(Q_{x}P_{y}-Q_{y}P_{x}\right)\hat{\mathbf{e}}_{z}$
is the phonon angular momentum and $\mathbf{B}_{\mathrm{eff}}=\left[0,0,B_{\mathrm{eff}}\right]$.
Solving the linearized equations of motion, we find the phonon eigenfrequencies
\begin{equation}
\Omega_{0,\pm}=\Omega_{0}\pm\kappa B_{\mathrm{eff}},\label{eq:_split_spectra}
\end{equation}
where the $\pm$-signs correspond to right- and left-handed motion,
respectively. 

\begin{figure}

\begin{centering}
\includegraphics[width=0.9\columnwidth]{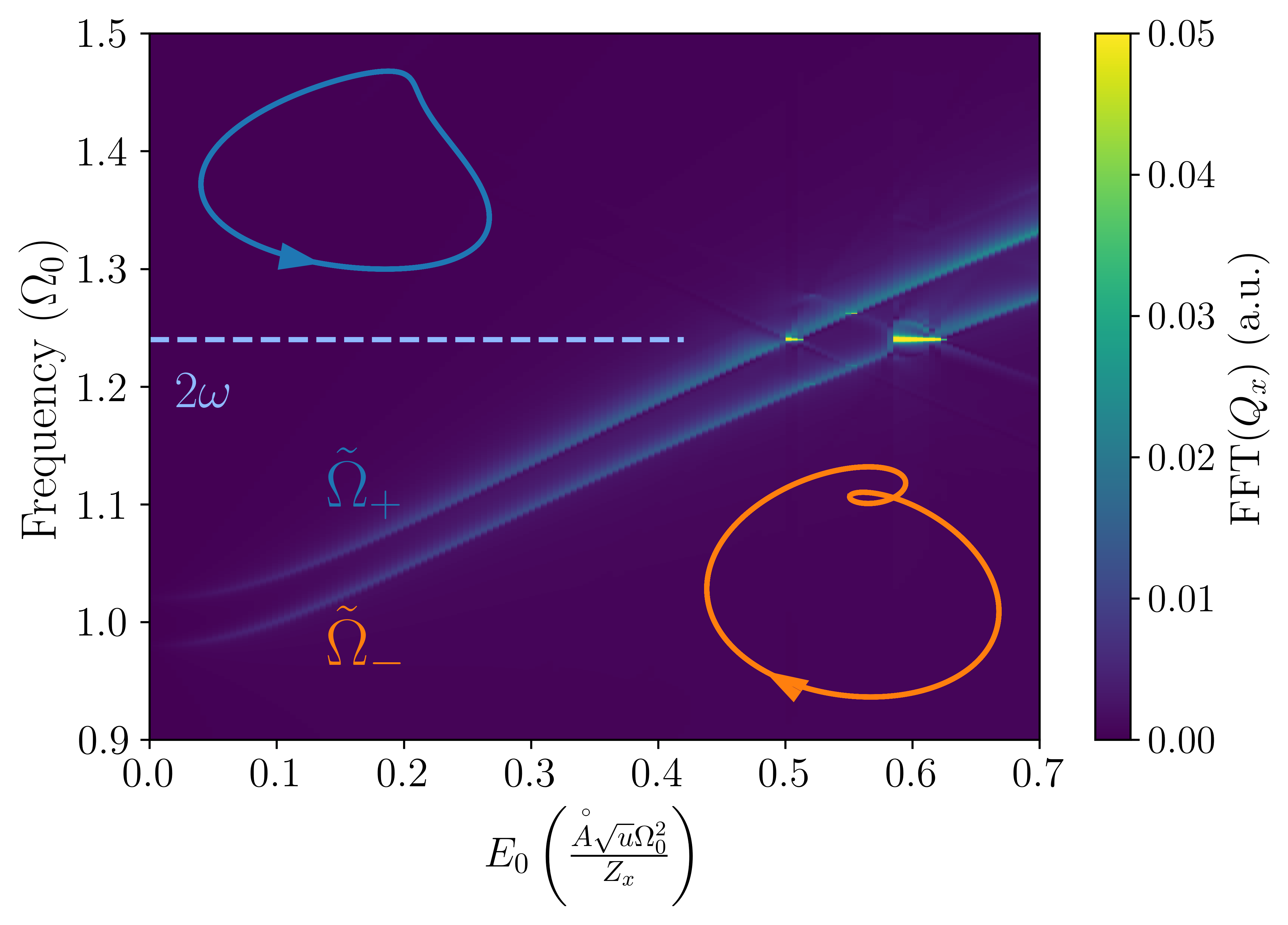}\caption{Symmetry breaking with non-degenerate chiral phonons {[}see Eq. (\ref{eq:_split_spectra}){]}.
The phonon frequencies are split according to Eq. (\ref{eq:_split_spectra}):
$\Omega_{\pm}$ corresponds to right/left-handed motion. The two modes
are accessed with light of opposite polarizations. To excite the $\Omega_{+}$
mode, we use $\mathbf{E}=E_{0}\left[-\left(1-\delta\right)\cos\omega t,\sin\omega t,0\right]$,
and for the $\Omega_{-}$ mode, $\mathbf{E}=E_{0}\left[\cos\omega t,\left(1-\delta\right)\sin\omega t,0\right]$,
with $\delta=0.25$ and $\omega=0.62\Omega$. As for degenerate chiral
phonons, a slight detuning from circularity $\delta$ is necessary,
in order to trigger the instability at half the resonance frequency
{[}see Eq. (\ref{eq:instability_Fx=00003DFy_expand}){]}. The figure
combines the results of two runs, in which the two chiralities were
simulated separately. The resonance frequencies exhibit a driving
amplitude dependent blue-shift, such that the instability occurs at
different powers, for the two chiralities.\label{fig:true_chirality} }
.
\par\end{centering}
\end{figure}

We find that the symmetry-breaking instability described above can
also be achieved with non-degenerate chiral phonons and present these
results in Fig. \ref{fig:true_chirality}.

\paragraph{Conclusion}

We have described a new, inversion-symmetry-breaking steady state
for driven nonlinear phonons. This state is characterized by strong
second harmonic generation and by the emergence of ferroelectric behavior.
These responses -- forbidden by the inversion symmetry of the underlying
lattice -- can serve as sharp experimental signatures of the physics
presented in this manuscript.

Ferroelectrics above the transition \citep{ginzburg1949theory_ferroelectrics,cochran1959_cochran_ferroelectrics1,cochran1960_cochran_ferroelectrics2}
are natural platforms for realizing the effects. The softening phonon
modes responsible for the transition will have low frequencies and
strong anharmonicity \citep{scott1974soft_modes_review,pal2021origin_thz_soft_modes,alyabyeva2021_barium_hexaferrite_soft_modes},
resulting in low threshold powers $E_{x,*}$ as shown in Eq. (\ref{eq:threshold_field}). 

The symmetry-breaking state can be induced with flat-top pulses as
short as five optical cycles, and at reachable field strengths (Table\ref{tab:Critical-driving-fields}).
Beyond possible applications for second harmonic generation, rectification,
and driven, on-demand ferroelectricity, the study of interactions
of phonons in the newly described state with electrons and collective
modes offers a route to novel out-of-equilibrium states.
\begin{acknowledgments}
We acknowledge comments by the anonymous referee of one of our previous
publications \citep{kiselev2019squid_spectroscopy}, which, eventually,
led to the research reported here, as well as useful discussions with
Mark Rudner. We also thank Jonas F. Karcher who motivated us to work
on this manuscript, by pointing out that ``a Nature article about
chiral phonons was on his newsfeed \citep{juraschek2025chiral_phonons_rev},
and chiral phonons seem to be a hot topic''. This project received
funding from the Horizon Europe Marie Sk\l odowska-Curie Action program
under Grant Agreement 101155351.
\end{acknowledgments}

\newpage
\section*{Appendix} 

\renewcommand{\thesection}{Appendix}%

\setcounter{figure}{0}
\renewcommand{\thefigure}{A\ \arabic{figure}}%

\setcounter{equation}{0}
\renewcommand{\theequation}{A\,\arabic{equation}}%

\subsection{Achieving symmetry breaking with pulsed excitations\label{subsec:Achieving-symmetry-breaking}}

\begin{figure}[h]

\begin{centering}
\includegraphics[width=0.75\columnwidth]{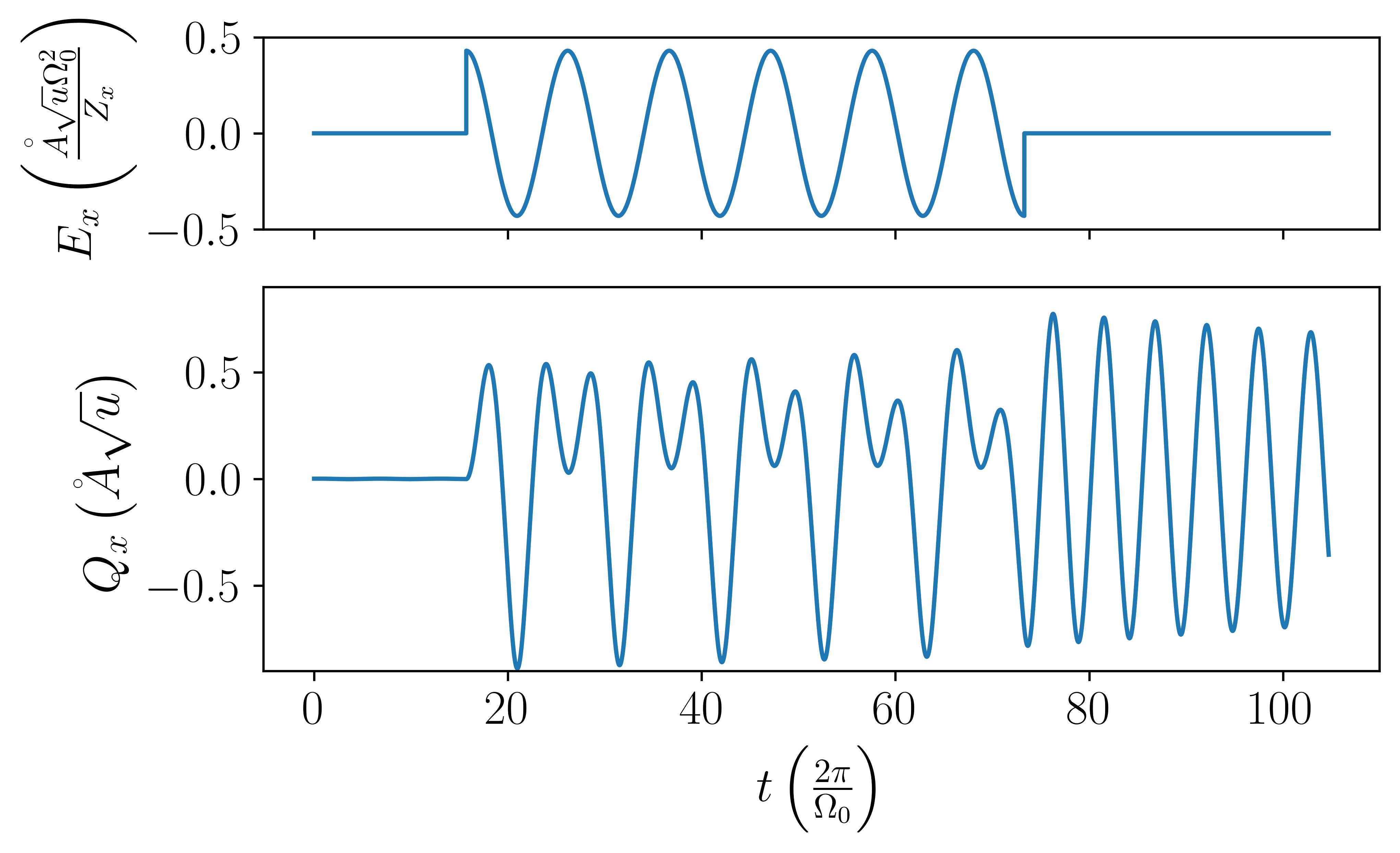}\caption{Reaching the symmetry-breaking state is possible with a short flat-top
pulse of only five optical cycles. The upper plot shows the driving
electric field, while the lower plot shows the response with visible
second harmonic and DC offset components. When the pulse ends, the
system switches to unforced oscillations at frequency $\Omega_{0}$.
The of $Q_{x}\left(t\right)$ is show in Fig.\ref{the_fig} of the
main text.\label{fig:pulses}}
\par\end{centering}
\end{figure}

We perform simulations to show the symmetry-breaking steady state
can be reached with short pulsed signals. We confine the oscillating
electric field in Eq. (\ref{eq:EoM_single_mode}) to a window of length
$T_{p}=5\cdot2\pi/\omega$. This mimics a flat-top pulse (see Fig.\ref{fig:pulses})
whose amplitude is constant throughout $T_{p}$. The constant amplitude
is necessary, because due to the blue-shift of $\tilde{\Omega}_{0}\left(E_{x}\right)$
both amplitude and frequency of the drive have to be matched with
each other (see Fig. \ref{the_fig}). Fig. \ref{the_fig} shows the
flat-top pulse and the response of the phonon mode $Q_{x}\left(t\right)$.

\subsubsection{A note on initial conditions and symmetry breaking\label{subsec:initial_conditions}}

We started our simulation runs at $t=0$ and used the forcing term
$E_{x}\left(t\right)=E_{0,x}\cos\left(\omega t\right)$, effectively
this corresponds to $E_{x}\left(t\right)\sim\Theta\left(t\right)\cos\left(\omega t\right)$.
The response of Eq. (\ref{eq:EoM_single_mode}) to this forcing necessarily
consists of particular solution oscillating at $\omega$ and persisting
for $t\rightarrow\infty$, and a homogeneous solution oscillating
at $\tilde{\Omega}_{0}$ that fixes the boundary conditions at $t=0$,
and decays due to damping. The homogeneous solution has both finite
$Q_{\mathrm{even}}$ and $Q_{\mathrm{odd}}$ components since $\tilde{\Omega}_{0}$
is not necessarily a multiple of $\omega$. It thus initializes the
parametric instability in Eq. (\ref{eq:prametric_osc_even}), which
gives $Q_{\mathrm{even}}\left(t\right)=0$ for $Q_{\mathrm{even}}\left(0\right)=0$
and $\dot{Q}_{\mathrm{even}}\left(0\right)=0$, acting as an initial
forcing for $Q_{\mathrm{even}}$, and determining the sign of the
symmetry breaking.

\subsection{Noise robustness\label{subsec:Noise-robustness}}

In our noise simulations, $\eta\left(t\right)$ is a normally-distributed
random function on a time grid with time points $t_{i}$ separated
by intervals $\Delta t=5\cdot10^{-3}/\Omega_{0}$. For simplicity,
we use cubic splines to interpolate $\eta\left(t\right)$ between
the $t_{i}$, and use a standard numerical integrator (SciPy's odeint).
This does not influence the physical results since $\Delta t\ll2\pi/\Omega_{0}$.
The symmetry-breaking steady state is stable up to a threshold noise
amplitude $A_{*}$. 

To relate the noise amplitude $A_{*}$ to a temperature, we calculate
the internal energy of the oscillator at finite temperatures via $U=-\partial_{\tilde{\beta}}\log Z$:
\begin{align*}
U & =-\frac{1}{2\pi\hbar}\frac{\partial}{\partial\left(1/k_{B}T\right)}\log\int dq\int dpe^{-\frac{H\left(p,q\right)}{k_{B}T}}\\
 & =\frac{1}{8}\left(\frac{\Omega^{4}K_{\frac{5}{4}}\left(\frac{\Omega^{4}}{8\beta k_{B}T}\right)}{\beta K_{\frac{1}{4}}\left(\frac{\Omega^{4}}{8\beta k_{B}T}\right)}-\frac{\Omega^{4}}{\beta}+2k_{B}T\right)\\
 & \approx k_{B}T-\frac{3\beta\left(k_{B}T\right)^{2}}{4\Omega^{4}}-\mathcal{O}\left(\beta^{2}\right).
\end{align*}
Assuming a phonon resonance frequency of $3\,$THz, and extracting
the internal energy $U$ corresponding to noise of amplitude $A_{*}$
(in absence of driving), we estimate that $A_{*}$ corresponds to
a temperature of about $700\,$K.

\subsection{Pumping via Raman-active modes\label{subsec:Auxiliary-resonances-and}}

In centrosymmetric materials, Raman- and IR-active modes are mutually
exclusive. They can be, however, nonlinearly coupled to each other
via the coupling Hamiltonian of Eq. (\ref{eq:Raman_coupling_H}),
where $Q_{R}$ is the Raman active and $Q_{x}$ the IR active mode.
The equations of motion for the coupled system can then be written
down as
\begin{align}
\ddot{Q}_{x}+2\gamma\dot{Q}_{x}+\Omega_{0}^{2}Q_{x}+\beta Q_{x}^{3} & =2AQ_{x}Q_{R}\nonumber \\
 & +Z_{x}E_{0}\cos\left(\omega t\right)
\end{align}
\begin{align}
 & \ddot{Q}_{R}+2\gamma\dot{Q}_{R}+\Omega_{R}^{2}Q_{R}+\beta_{R,2}Q_{R}^{2}+\beta_{R,3}Q_{R}^{3}.\nonumber \\
 & \qquad=AQ_{x}^{2}+\frac{1}{2}\chi_{R}E_{0}^{2}\left(1+\cos\left(2\omega t\right)\right)
\end{align}
We assume that the drive is in resonance with the Raman-active mode,
such that $2\omega\approx\Omega_{R}$. Furthermore, we assume that
$\Omega_{0}$ is distinct from $\Omega_{R}$, such that the driving
is off-resonant with $\Omega_{0}$, and the direct, off-resonant driving
of $Q_{x}$ via $Z_{x}$ is neglectable. We will furthermore be interested
in instabilites of $Q_{x}$ induced by the Raman mode. At the instability
onset, which we want to study here, $Q_{x}$ will be small, and we
can neglect higher order terms in this variable. The simplified system
of equations reads
\begin{align}
\ddot{Q}_{x}+2\gamma\dot{Q}_{x}+\Omega_{0}^{2}Q_{x}-2AQ_{x}Q_{R} & =0
\end{align}
\begin{align}
 & \ddot{Q}_{R}+2\gamma\dot{Q}_{R}+\Omega_{R}^{2}Q_{R}+\beta_{R,2}Q_{R}^{2}+\beta_{R,3}Q_{R}^{3}\nonumber \\
 & \qquad=\frac{1}{2}\chi_{R}E_{0}^{2}\left(1+\cos\left(2\omega t\right)\right).
\end{align}
Inverting the second equation, we find, neglecting higher harmonics
\begin{equation}
Q_{R}\approx F_{R,0}\left(E_{0}\right)+F_{R,2}\left(E_{0}\right)\cos\left(2\omega t\right),
\end{equation}
where $F_{R,0}$ and $F_{R,2}$ are the amplitudes of DC and second
harmonic (with respect to the driving frequency) components. Inserting
this into the first equation gives
\begin{equation}
\ddot{Q}_{x}+2\gamma\dot{Q}_{x}+\tilde{\Omega}_{0}\left(E_{0}\right)\left[1-h\left(E_{0}\right)\cos\left(2\omega t\right)\right]Q_{x}=0.
\end{equation}
Here $\tilde{\Omega}_{0}\left(E_{0}\right)=\Omega_{0}^{2}+F_{R,0}\left(E_{0}\right)$
is the effective, blue shifted resonance frequency of the IR active
mode and $h\left(E_{0}\right)=2AF_{R,2}\left(E_{0}\right)$. This
equation maps onto Eq. () of the main text. Showing, that a Raman-active
mode can be used to trigger the symmetry-breaking instability of the
IR active mode. The frequency of the resonant Raman-active mode $\Omega_{R}\approx2\omega$
has to be reasonbly close to $\Omega_{0}$. However, some detuning
is possible, because the effective resonance frequency $\tilde{\Omega}_{0}\left(E_{0}\right)$
experiences a blue shift. This blue shift can be exploted to adjust
the effective resonance frequency by adjusting the driving amplitude
$E_{0}$.

Similarly, an auxiliary IR-active mode with a frequency close to half
the resonance $\Omega_{0}$ can be used, instead of the Raman-active
mode. An advantage of using an auxiliary (Raman or IR-active mode)
is that the mode will story the energy supplied by the drive over
many cycles. Thus the threshold amplitude for the instability might
be easier reach.

\begin{figure}

\begin{centering}
\includegraphics[width=0.9\columnwidth]{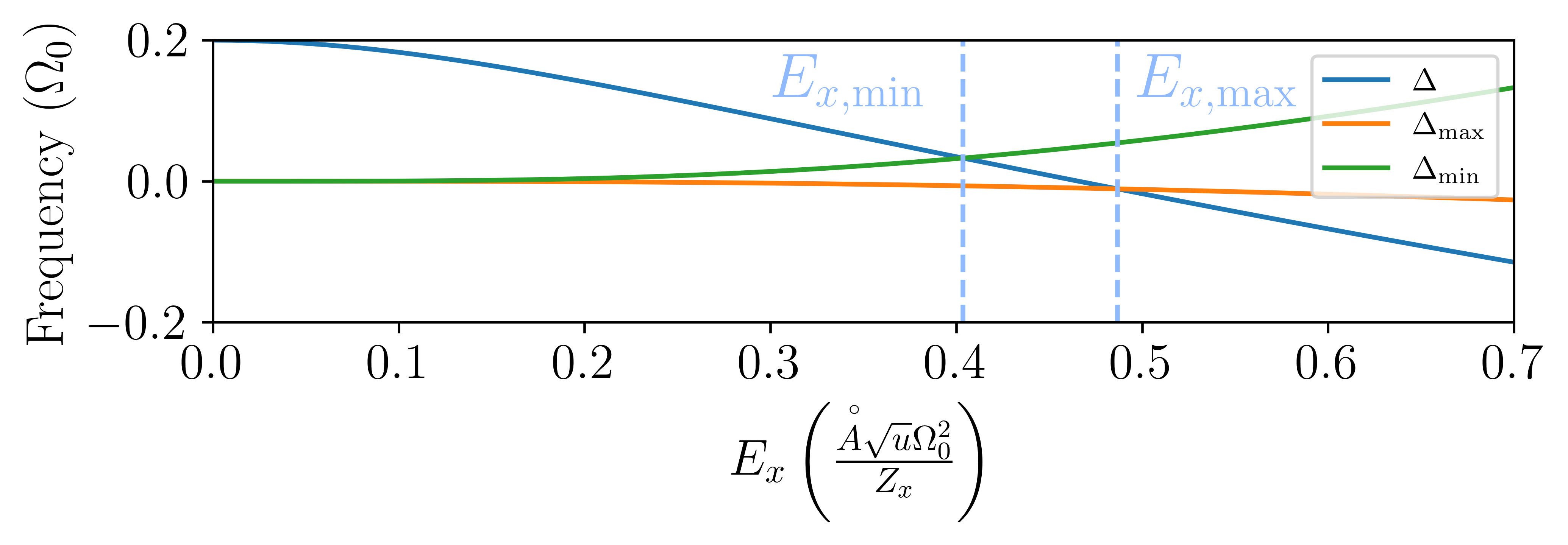}\caption{Illustration of the instability condition of Eq. (\ref{eq:instability_interval})
for the same parameteres as in Fig. \ref{the_fig}a).}
\par\end{centering}
\end{figure}

\subsection{Collective instability of the x and y modes\label{subsec:Collective-instability-of}}

The equations of motion (with the addition of a damping term) following
form Eq. (\ref{eq:Phonon_H}) read

\begin{align}
\ddot{Q}_{x}+2\gamma\dot{Q}_{x}+\Omega_{0}^{2}Q_{x}+\beta Q_{x}\left(Q_{x}^{2}+Q_{y}^{2}\right) & =Z_{x}E_{x}\cos\omega t\nonumber \\
\ddot{Q}_{y}+2\gamma\dot{Q}_{y}+\Omega_{0}^{2}Q_{y}+\beta Q_{y}\left(Q_{x}^{2}+Q_{y}^{2}\right) & =Z_{y}E_{y}\sin\omega t.\label{eq:Eqs_of_motion_circular_1}
\end{align}
We first derive the two-component analogue of Eq. (\ref{eq:prametric_osc_even}),
which is given by
\begin{align}
 & \ddot{Q}_{x/y,\mathrm{even}}+2\gamma\dot{Q}_{x/y,\mathrm{even}}\nonumber \\
 & \qquad+\left[\tilde{\Omega}_{x/y}^{2}\pm\frac{\alpha}{2}\left(3F_{x/y,1}^{2}-F_{y/x,1}^{2}\right)\cos\left(2\omega t\right)\right]Q_{x/y,\mathrm{even}}\nonumber \\
 & \qquad+\alpha F_{x/y,1}F_{y/x,1}\sin\left(2\omega t\right)Q_{y/x,\mathrm{even}}=0\label{eq:param_even}
\end{align}
with $\tilde{\Omega}_{x/y}^{2}=\Omega_{0}^{2}+\alpha\left(3F_{x/y}^{2}+F_{y/x}^{2}\right)/2$,
where $F_{y}$ is defined analogously to $F_{x}$ in Eq. (\ref{eq:duffing_1harm_response}).
As above, we expect parametric resonances near $2\omega=\tilde{\Omega}_{x/y}$,
where $\tilde{\Omega}_{x}\neq\tilde{\Omega}_{y}$ for $F_{x}\neq F_{y}$.
For now, let us choose the case $2\omega=\tilde{\Omega}_{x}$, such
that the instability occurs for the $Q_{x}\left(t\right)$ component. 

The oscillating terms in Eqs. (\ref{eq:param_even}) couple harmonics
with frequencies $2\omega$, $4\omega$, ... and DC terms. For the
stability analysis, we therefore choose the ansatz
\begin{align}
Q_{i,\mathrm{even}} & =a_{i,1}\sin\left(2\omega t\right)+a_{i,2}\sin\left(4\omega t\right)+b_{i,0}\nonumber \\
 & +b_{i,1}\cos\left(2\omega t\right)+b_{i,2}\cos\left(4\omega t\right).\label{eq:4_freq_ansatz}
\end{align}
Furthermore, we neglect $\gamma$ for the duration of this analysis.
While $\gamma$ determines the instability threshold amplitudes of
the electromagnetic fields {[}see Eq. (\ref{eq:threshold_field}){]},
its effects become less important for driving amplitudes above the
threshold, i.e., for any driving amplitude, $\gamma$ can be always
chosen small enough that our analysis is accurate. We again search
for the instability window for the detuning $\Delta=2\omega-\tilde{\Omega}_{x}$,
such that the mode amplitudes $a_{i,n}$ and $b_{i,n}$ grow exponentially
for $\Delta_{\mathrm{min}}<\Delta<\Delta_{\mathrm{max}}$. At $\Delta=\Delta_{\mathrm{max}/\mathrm{min}}$,
the amplitudes will be constant. The boundaries of the instability
interval $\Delta_{\mathrm{max}/\mathrm{min}}$ are then found by inserting
the ansatz (\ref{eq:4_freq_ansatz}) into Eqs. (\ref{eq:param_even})
and assuming that $a_{i,n}$ and $b_{i,n}$ are indeed constant \citep{Landau_Lifshitz_Mechanics}.
After a lengthy calculation, in which we compare the coefficients
of different harmonics after inserting the ansatz (\ref{eq:4_freq_ansatz})
into Eqs. (\ref{eq:param_even}), we find that, to fourth order in
$F_{x,1}$ and $F_{y,1}$
\begin{align}
\Delta_{\mathrm{max}}-\Delta_{\mathrm{min}} & =F_{y,1}^{4}\left(\frac{\beta^{2}}{16\Omega_{0}^{3}}+\frac{287\beta^{4}F_{x,1}^{4}}{576\Omega_{0}^{7}}+\frac{47\beta^{3}F_{x,1}^{2}}{192\Omega_{0}^{5}}\right)\nonumber \\
 & +F_{y,1}^{2}\left(\frac{77\beta^{3}F_{x,1}^{4}}{192\Omega_{0}^{5}}-\frac{5\beta^{2}F_{x,1}^{2}}{8\Omega_{0}^{3}}\right)+\frac{9\beta^{2}F_{x,1}^{4}}{16\Omega_{0}^{3}}.\label{eq:instability_Fx_Fy_expand}
\end{align}
The full result is too long to be quoted here but is easily found
using computer algebra. Eq. (\ref{eq:instability_Fx_Fy_expand}) is
valid for small $F_{x}$ and $F_{y}$, i.e. for small driving amplitudes.
It is interesting to study the behavior of $\Delta$ close to $F_{y,1}=F_{x,1}$,
i.e. for nearly perfect circular polarization. Writing $F_{y,1}=F_{x,1}+F_{\epsilon}$,
we find
\begin{align}
\Delta_{\mathrm{max}}-\Delta_{\mathrm{min}} & =-F_{\epsilon}^{3}\left(\frac{9\Omega_{0}}{8F_{x,1}^{3}}+\frac{\beta^{2}F_{x,1}}{\Omega_{0}^{3}}+\frac{51\beta}{16F_{x,1}\Omega_{0}}\right).\label{eq:instability_Fx=00003DFy_expand}
\end{align}
Notice that for $F_{y,1}\geq F_{x,1}$ (we choose both amplitudes
positive w.l.o.g.), we have $\Delta_{\mathrm{max}}\leq\Delta_{\mathrm{min}}$,
which indicates that the system is stable. For $F_{y,1}>F_{x,1}$,
the $Q_{x}$ and $Q_{y}$ components switch places, and the instability
occurs for $2\omega=\tilde{\Omega}_{y}$. The analysis for this case
is completely analogous with $F_{x,1}$ and $F_{y,1}$ , as well as
$\Omega_{x}$ and $\Omega_{y}$ interchanged. We therefore conclude
that, in general, the instability occurs either for the $Q_{x}$ or
the $Q_{y}$ component, depending on whether $F_{x,1}$ or $F_{y,1}$
is larger. For $E_{x}=E_{y}$, resulting in $F_{x}=F_{y}$, we find
$\Delta_{\mathrm{min}}=\Delta_{\mathrm{max}}$ and the instability
cannot be reached. \citep{Note2}.


\begin{thebibliography}{10}

\bibitem{basov2017_on_demand}
D.~Basov, R.~Averitt, and D.~Hsieh, \textit{Towards properties on demand in
  quantum materials}, Nature materials \textbf{16}, 1077 (2017).

\bibitem{bloch2022strongly_corr_el_photon}
J.~Bloch, A.~Cavalleri, V.~Galitski, M.~Hafezi, and A.~Rubio, \textit{Strongly
  correlated electron--photon systems}, Nature \textbf{606}, 41 (2022).

\bibitem{rudner2020band}
M.~S. Rudner and N.~H. Lindner, \textit{Band structure engineering and
  non-equilibrium dynamics in floquet topological insulators}, Nature reviews
  physics \textbf{2}, 229 (2020).

\bibitem{forst2011nonlinear_phononics}
M.~F{\"o}rst, C.~Manzoni, S.~Kaiser, Y.~Tomioka, Y.~Tokura, R.~Merlin, and
  A.~Cavalleri, \textit{Nonlinear phononics as an ultrafast route to lattice
  control}, Nature Physics \textbf{7}, 854 (2011).

\bibitem{mankowsky2016nonlinear_phononics_2}
R.~Mankowsky, M.~F{\"o}rst, and A.~Cavalleri, \textit{Non-equilibrium control
  of complex solids by nonlinear phononics}, Reports on Progress in Physics
  \textbf{79}, 064503 (2016).

\bibitem{mankowsky2014_enhanced_superconductivity}
R.~Mankowsky, A.~Subedi, M.~F{\"o}rst, S.~O. Mariager, M.~Chollet, H.~Lemke,
  J.~S. Robinson, J.~M. Glownia, M.~P. Minitti, A.~Frano \textit{et~al.},
  \textit{Nonlinear lattice dynamics as a basis for enhanced superconductivity
  in yba2cu3o6. 5}, Nature \textbf{516}, 71 (2014).

\bibitem{knap2016phonon_superconduct_theory}
M.~Knap, M.~Babadi, G.~Refael, I.~Martin, and E.~Demler, \textit{Dynamical
  cooper pairing in nonequilibrium electron-phonon systems}, Physical Review B
  \textbf{94}, 214504 (2016).

\bibitem{babadi2017phonon_superconduct_theory}
M.~Babadi, M.~Knap, I.~Martin, G.~Refael, and E.~Demler, \textit{Theory of
  parametrically amplified electron-phonon superconductivity}, Physical Review
  B \textbf{96}, 014512 (2017).

\bibitem{cavalleri2018photo_induced_supercond}
A.~Cavalleri, \textit{Photo-induced superconductivity}, Contemporary Physics
  \textbf{59}, 31 (2018).

\bibitem{liu2020_res_phonon_light_induced_superc}
B.~Liu, M.~F{\"o}rst, M.~Fechner, D.~Nicoletti, J.~Porras, T.~Loew, B.~Keimer,
  and A.~Cavalleri, \textit{Pump frequency resonances for light-induced
  incipient superconductivity in $yba_2cu_3o_{6.5}$}, Physical Review X
  \textbf{10}, 011053 (2020).

\bibitem{fechner2018magnetophononics_review}
M.~Fechner, A.~Sukhov, L.~Chotorlishvili, C.~Kenel, J.~Berakdar, and
  N.~Spaldin, \textit{Magnetophononics: Ultrafast spin control through the
  lattice}, Physical review materials \textbf{2}, 064401 (2018).

\bibitem{afanasiev2021ultrafast_control_magnetic_phonons}
D.~Afanasiev, J.~Hortensius, B.~Ivanov, A.~Sasani, E.~Bousquet, Y.~Blanter,
  R.~Mikhaylovskiy, A.~Kimel, and A.~Caviglia, \textit{Ultrafast control of
  magnetic interactions via light-driven phonons}, Nature materials
  \textbf{20}, 607 (2021).

\bibitem{disa2023_phonon_ferromag_induce}
A.~Disa, J.~Curtis, M.~Fechner, A.~Liu, A.~Von~Hoegen, M.~F{\"o}rst, T.~Nova,
  P.~Narang, A.~Maljuk, A.~Boris \textit{et~al.}, \textit{Photo-induced
  high-temperature ferromagnetism in ytio3}, Nature \textbf{617}, 73 (2023).

\bibitem{luo2025terahertz_control_magno_phononics}
T.~Luo, H.~Ning, B.~Ilyas, A.~von Hoegen, E.~Vi{\~n}as~Bostr{\"o}m, J.~Park,
  J.~Kim, J.-G. Park, D.~M. Juraschek, A.~Rubio \textit{et~al.},
  \textit{Terahertz control of linear and nonlinear magno-phononics}, Nature
  Communications \textbf{16}, 6863 (2025).

\bibitem{nova2019phonon_ferroelectricity}
T.~Nova, A.~Disa, M.~Fechner, and A.~Cavalleri, \textit{Metastable
  ferroelectricity in optically strained srtio3}, Science \textbf{364}, 1075
  (2019).

\bibitem{ning2023_phonon_induced_hidden_quadrupolar}
H.~Ning, O.~Mehio, X.~Li, M.~Buchhold, M.~Driesse, H.~Zhao, G.~Cao, and
  D.~Hsieh, \textit{A coherent phonon-induced hidden quadrupolar ordered state
  in ca2ruo4}, Nature Communications \textbf{14}, 8258 (2023).

\bibitem{kaplan2025_spatiotemporal_phonons}
D.~Kaplan, P.~A. Volkov, A.~Chakraborty, Z.~Zhuang, and P.~Chandra,
  \textit{Tunable spatiotemporal orders in driven insulators}, Physical review
  letters \textbf{134}, 066902 (2025).

\bibitem{kaplan2025spatiotemporal_phonons_first_principles}
D.~Kaplan, P.~A. Volkov, J.~Coulter, S.~Zhang, and P.~Chandra,
  \textit{Spatiotemporal order and parametric instabilities from
  first-principles}, arXiv preprint arXiv:2507.14110  (2025).

\bibitem{kiselev2024MFPD_plasmons_nat}
E.~I. Kiselev, M.~S. Rudner, and N.~H. Lindner, \textit{Inducing exceptional
  points, enhancing plasmon quality and creating correlated plasmon states with
  modulated floquet parametric driving}, Nature Communications \textbf{15},
  9914 (2024).

\bibitem{kiselev2024_PRB_MFPD}
E.~I. Kiselev, Y.~Pan, and N.~H. Lindner, \textit{Light-controlled terahertz
  plasmonic time-varying media: Momentum gaps, entangled plasmon pairs, and
  pulse-induced time reversal}, Physical Review B \textbf{110}, L241411 (2024).

\bibitem{kiselev2025exciting}
E.~I. Kiselev, J.~F. Karcher, M.~S. Rudner, R.~Duine, and N.~H. Lindner,
  \textit{Exciting terahertz magnons with amplitude modulated light: spin
  pumping, squeezed states, symmetry breaking and pattern formation}, arXiv
  preprint arXiv:2507.08147  (2025).

\bibitem{wanic2025entanglement}
M.~Wanic, C.~Jasiukiewicz, Z.~Toklikishvili, V.~Jandieri, M.~Trybus,
  E.~Jartych, S.~Mishra, and L.~Chotorlishvili, \textit{Entanglement properties
  of photon--magnon crystal from nonlinear perspective}, Physica D: Nonlinear
  Phenomena \textbf{476}, 134699 (2025).

\bibitem{kaplan2025optically_induced_goldstone_faraday}
D.~Kaplan, P.~A. Volkov, A.~Cavalleri, and P.~Chandra,
  \textit{Optically-induced faraday-goldstone waves}, arXiv preprint
  arXiv:2511.07320  (2025).

\bibitem{wanic2024magnetoelectric}
M.~Wanic, Z.~Toklikishvili, S.~Mishra, M.~Trybus, and L.~Chotorlishvili,
  \textit{Magnetoelectric fractals, magnetoelectric parametric resonance and
  hopf bifurcation}, Physica D: Nonlinear Phenomena \textbf{467}, 134257
  (2024).

\bibitem{cartella2018parametric}
A.~Cartella, T.~F. Nova, M.~Fechner, R.~Merlin, and A.~Cavalleri,
  \textit{Parametric amplification of optical phonons}, Proceedings of the
  National Academy of Sciences \textbf{115}, 12148 (2018).

\bibitem{buzzi2021higgs_lasing}
M.~Buzzi, G.~Jotzu, A.~Cavalleri, J.~I. Cirac, E.~A. Demler, B.~I. Halperin,
  M.~D. Lukin, T.~Shi, Y.~Wang, and D.~Podolsky, \textit{Higgs-mediated optical
  amplification in a nonequilibrium superconductor}, Physical Review X
  \textbf{11}, 011055 (2021).

\bibitem{michael2024photonic}
M.~H. Michael, S.~R.~U. Haque, L.~Windgaetter, S.~Latini, Y.~Zhang, A.~Rubio,
  R.~D. Averitt, and E.~Demler, \textit{Photonic time-crystalline behaviour
  mediated by phonon squeezing in t a 2 n i s e 5}, Nature Communications
  \textbf{15}, 3638 (2024).

\bibitem{Boyd_nonlinear_optics}
R.~W. Boyd, \textit{Nonlinear Optics}, Academic press (2008). ISBN
  978-0123694706.

\bibitem{li2019_driven_metastable_ferroelectricity}
X.~Li, T.~Qiu, J.~Zhang, E.~Baldini, J.~Lu, A.~M. Rappe, and K.~A. Nelson,
  \textit{Terahertz field--induced ferroelectricity in quantum paraelectric
  srtio3}, Science \textbf{364}, 1079 (2019).

\bibitem{disa2020polarizing_symmetry_breaking_light_nonlinearity}
A.~S. Disa, M.~Fechner, T.~F. Nova, B.~Liu, M.~F{\"o}rst, D.~Prabhakaran, P.~G.
  Radaelli, and A.~Cavalleri, \textit{Polarizing an antiferromagnet by optical
  engineering of the crystal field}, Nature Physics \textbf{16}, 937 (2020).

\bibitem{juraschek2025chiral_phonons_rev}
D.~M. Juraschek, R.~M. Geilhufe, H.~Zhu, M.~Basini, P.~Baum, A.~Baydin,
  S.~Chaudhary, M.~Fechner, B.~Flebus, G.~Grissonnanche \textit{et~al.},
  \textit{Chiral phonons}, Nature Physics (1--9) (2025).

\bibitem{kahana2024_chiral_rectification}
T.~Kahana, D.~A. Bustamante~Lopez, and D.~M. Juraschek, \textit{Light-induced
  magnetization from magnonic rectification}, Science Advances \textbf{10},
  eado0722 (2024).

\bibitem{luo2023large_magnetic_chiral_phonons}
J.~Luo, T.~Lin, J.~Zhang, X.~Chen, E.~R. Blackert, R.~Xu, B.~I. Yakobson, and
  H.~Zhu, \textit{Large effective magnetic fields from chiral phonons in
  rare-earth halides}, Science \textbf{382}, 698 (2023).

\bibitem{juraschek2022giant_magn_field_chiral_phonons}
D.~M. Juraschek, T.~Neuman, and P.~Narang, \textit{Giant effective magnetic
  fields from optically driven chiral phonons in 4 f paramagnets}, Physical
  Review Research \textbf{4}, 013129 (2022).

\bibitem{xiong2022effective_magn_field_chiral}
G.~Xiong, H.~Chen, D.~Ma, and L.~Zhang, \textit{Effective magnetic fields
  induced by chiral phonons}, Physical Review B \textbf{106}, 144302 (2022).

\bibitem{yaniv2025multicolor}
O.~Yaniv and D.~M. Juraschek, \textit{Multicolor phonon excitation in terahertz
  cavities}, Physical Review Letters \textbf{135}, 246901 (2025).

\bibitem{Landau_Lifshitz_Mechanics}
L.~D. Landau and E.~M. Lifshitz, \textit{Mechanics}, Butterworth-Heinemann
  (1976). ISBN 978-0750628969.

\bibitem{turyn1993damped_Mathieu}
L.~Turyn, \textit{The damped mathieu equation}, Quarterly of applied
  mathematics \textbf{51}, 389 (1993).

\bibitem{Note1}
The equations on p. 390 of this reference contain a typo.

\bibitem{vonHoegen2018probing}
A.~von Hoegen, R.~Mankowsky, M.~Fechner, M.~F{\"o}rst, and A.~Cavalleri,
  \textit{Probing the interatomic potential of solids with strong-field
  nonlinear phononics}, Nature \textbf{555}, 79 (2018).

\bibitem{ys1991_duffing_symm_breaking_SHG}
C.~Olson and M.~Olsson, \textit{Dynamical symmetry breaking and chaos in
  duffing's equation}, Am. J. Phys \textbf{59} (1991).

\bibitem{xu2016pinched_hysteresis_ferroelectrics}
B.~Xu, C.~Paillard, B.~Dkhil, and L.~Bellaiche, \textit{Pinched hysteresis loop
  in defect-free ferroelectric materials}, Physical Review B \textbf{94},
  140101 (2016).

\bibitem{cheng2020_dirac_semi_phonon_zeeman}
B.~Cheng, T.~Schumann, Y.~Wang, X.~Zhang, D.~Barbalas, S.~Stemmer, and
  N.~Armitage, \textit{A large effective phonon magnetic moment in a dirac
  semimetal}, Nano letters \textbf{20}, 5991 (2020).

\bibitem{mustafa2025origin_phonon_zeeman_mos2}
H.~Mustafa, C.~Nnokwe, G.~Ye, M.~Fang, S.~Chaudhary, J.-A. Yan, K.~Wu, C.~J.
  Cunningham, C.~M. Hemesath, A.~J. Stollenwerk \textit{et~al.}, \textit{Origin
  of large effective phonon magnetic moments in monolayer mos2}, ACS nano
  \textbf{19}, 11241 (2025).

\bibitem{zhang2015chiral_phonons_hexagonal}
L.~Zhang and Q.~Niu, \textit{Chiral phonons at high-symmetry points in
  monolayer hexagonal lattices}, Physical review letters \textbf{115}, 115502
  (2015).

\bibitem{zhu2018observation_chiral_phonons}
H.~Zhu, J.~Yi, M.-Y. Li, J.~Xiao, L.~Zhang, C.-W. Yang, R.~A. Kaindl, L.-J. Li,
  Y.~Wang, and X.~Zhang, \textit{Observation of chiral phonons}, Science
  \textbf{359}, 579 (2018).

\bibitem{ishito2023truly_chiral_phonons}
K.~Ishito, H.~Mao, Y.~Kousaka, Y.~Togawa, S.~Iwasaki, T.~Zhang, S.~Murakami,
  J.-i. Kishine, and T.~Satoh, \textit{Truly chiral phonons in $\alpha$-hgs},
  Nature Physics \textbf{19}, 35 (2023).

\bibitem{ueda2023chiral_phonons_xrays}
H.~Ueda, M.~Garcia-Fernandez, S.~Agrestini, C.~P. Romao, J.~van~den Brink,
  N.~A. Spaldin, K.-J. Zhou, and U.~Staub, \textit{Chiral phonons in quartz
  probed by x-rays}, Nature \textbf{618}, 946 (2023).

\bibitem{bonini2023CrI3_chiral_phonons}
J.~Bonini, S.~Ren, D.~Vanderbilt, M.~Stengel, C.~E. Dreyer, and S.~Coh,
  \textit{Frequency splitting of chiral phonons from broken time-reversal
  symmetry in cri 3}, Physical review letters \textbf{130}, 086701 (2023).

\bibitem{chen2019chiral_chen_review}
H.~Chen, W.~Zhang, Q.~Niu, and L.~Zhang, \textit{Chiral phonons in
  two-dimensional materials}, 2D Materials \textbf{6}, 012002 (2019).

\bibitem{ginzburg1949theory_ferroelectrics}
V.~Ginzburg, \textit{Theory of ferroelectric phenomena}, Usp. Fiz. Nauk
  \textbf{38}, 490 (1949).

\bibitem{cochran1959_cochran_ferroelectrics1}
W.~Cochran, \textit{Crystal stability and the theory of ferroelectricity},
  Physical Review Letters \textbf{3}, 412 (1959).

\bibitem{cochran1960_cochran_ferroelectrics2}
W.~Cochran, \textit{Crystal stability and the theory of ferroelectricity},
  Advances in Physics \textbf{9}, 387 (1960).

\bibitem{scott1974soft_modes_review}
J.~Scott, \textit{Soft-mode spectroscopy: Experimental studies of structural
  phase transitions}, Reviews of Modern Physics \textbf{46}, 83 (1974).

\bibitem{pal2021origin_thz_soft_modes}
S.~Pal, N.~Strkalj, C.-J. Yang, M.~C. Weber, M.~Trassin, M.~Woerner, and
  M.~Fiebig, \textit{Origin of terahertz soft-mode nonlinearities in
  ferroelectric perovskites}, Physical Review X \textbf{11}, 021023 (2021).

\bibitem{alyabyeva2021_barium_hexaferrite_soft_modes}
L.~N. Alyabyeva, A.~S. Prokhorov, D.~Vinnik, V.~B. Anzin, A.~Ahmed,
  A.~Mikheykin, P.~Bednyakov, C.~Kadlec, F.~Kadlec, E.~de~Prado
  \textit{et~al.}, \textit{Lead-substituted barium hexaferrite for tunable
  terahertz optoelectronics}, NPG Asia Materials \textbf{13}, 63 (2021).

\bibitem{kiselev2019squid_spectroscopy}
E.~Kiselev, A.~Averkin, M.~Fistul, V.~Koshelets, and A.~Ustinov,
  \textit{Two-tone spectroscopy of a squid metamaterial in the nonlinear
  regime}, Physical Review Research \textbf{1}, 033096 (2019).

\bibitem{Note2}
We note, that this must not necessarily be the case for other, less generic types of nonlinearities than the one considered here.

\end{thebibliography}
\end{document}